# Inducing and Mitigating a Self-Reinforcing Degradation in Decision-making Teams

Paul Hubbard, Alexander Kott, Michael Martin

Computational analyses of organizational behavior employ either intellective or emulative models [1]. The model described in this paper[1] is an intellective models. Intellective models are usually abstract, small, relatively simple, and incorporate only a few parameters. They help modelers make general predictions about trends and the relative benefit of organizational changes, identify the range of likely behaviors, and qualitatively compare the impact of different types of policies or technologies on expected behaviors. Emulative models are much more detailed, incorporate a large number of parameters or rules, are more difficult and expensive to construct, and require large volumes of input data. They help modelers make specific predictions about quantitative characteristics of organizational behavior for specific organizations under specific conditions, and compare quantitatively the expected impact of alternative organizational designs or policy changes.

The models in this paper demonstrate how self-reinforcing error due to positive feedback can lead to overload and saturation of decision-making elements, and ultimately the cascading collapse of an organization due to the propagation of overload and erroneous decisions throughout the organization.

We begin the paper with an analysis of the stability of the decision-making aspects of command organizations from a system-theoretic perspective. A simple dynamic model shows how an organization can enter into a self-reinforcing cycle of increasing decision workload until the demand for decisions exceeds the decision-making capacity of the organization. In this model we consider only two components corresponding to two layers of an organization—an upper command layer and a subordinate execution layer. We show that even this simple model offers useful insights into conditions under which an organization can experience a rapid decrease in decision quality.

---

[1] This paper is based in part on the paper *Instability of Distributed Decision Making in Command and Control Systems*, by Kott, A., Hubbard P. and Martin, M., which appeared in Proceedings of the 2001 American Control Conference, Arlington, VA, June 2001, © 2001 IEEE. It is also based in part on the paper *Managing Responsibility and Information Flow in Dynamic Team Decision-Making*, by Hubbard, P., Kott, A., and Martin, M., published in the Proceedings of Command and Control Research and Technology Symposium, 2002.

We then extend the model to more complex networked organizations and show that they also experience a form of self-reinforcing degradation. In particular, we find that the degradation in decision quality has a tendency to propagate through the hierarchical structure, i.e. overload at one location affects other locations by overloading the higher-level components which then in turn overload their subordinates.

But how would one devise measures that actually induce such a malfunction in an enemy organization? Conversely, how would one devise a set of actions that mitigate a malfunction in their own organization? Our computational experiments suggest several strategies for mitigating this type of malfunction: dumping excessive load, empowering lower echelons, minimizing the need for coordination, using command-by-negation, insulating weak performers, and applying on-line diagnostics. Further, a suitable compensating component, e.g. a brokering mechanism that dynamically re-distributes responsibilities within the organization as it begins to malfunction, can dramatically increase the envelope of stable performance. We describe a method to allocate decision responsibility and arrange information flow dynamically within a team of decision-makers for command and control. We argue that dynamic modification of the decision responsibilities and information-sharing links within a decision-making team can either degrade or improve (depending on the intent of the one who performs the modifications) stability and performance in terms of quality of decisions produced by the team.

*A Simple Model for Self-Reinforcing Decision Overload*

Let us consider a network of decision-making entities, perhaps individuals, teams of individuals, information-processing tools, or artificial agents, operating jointly in accordance with organizational procedures and protocols. Such a decision-making organization acquires, transforms, generates and disseminates information in order to acquire, allocate, and deploy its resources so that its objectives can be accomplished efficiently and effectively. A decision-making organization's ultimate product is the commands it issues to those operational elements that execute direct effects on the environment of the organization. We label the totality of these executing elements as the *field*. The field may include salespeople who are trying to affect the behavior of the buyers in the market; workers who assemble the products; trading floor clerks who execute the transactions; pilots of military aircraft who fly to bomb their targets. To illustrate the kinds of malfunctions we wish to model here, consider the following scenarios.

Scenario A. *After a major financial loss, a corporation forces one of its underperforming divisions into a major restructuring. Several new senior managers and advisors are brought into the divisional operations. The existing personnel dedicate a large fraction of their time to explaining and justifying their decisions to the new managers, modifying their procedures and plans according to the new guidance. The day-to-day decisions receive less attention and their quality suffers. Mistakes are made more often. The field personnel resent the erroneous guidance, and morale and discipline decline. Performance of the division suffers even further. The corporate management decides to step up the restructuring…and the vicious cycle continues.*

Scenario B. *A corporation faces a new, unexpected tactic employed by its competitor. The tactic is successful and rapidly makes the business plans and procedures of the corporation inapplicable. Management attempts to introduce new ideas and approaches. Field personnel are bewildered and call for explanations and support. Decisions with new unfamiliar approaches become harder just as the attention of management is distracted by the competitor's new tactics. The quality of decisions deteriorates. Management's confidence plummets and decisions take even more effort. The competition exploits the errors and continues to succeed in altering the market position, which in turn requires more adjustments in corporate business tactic… which in turn causes more confusion and errors…*

Qualitative discussions of challenges and phenomena in decision-making organizations are numerous, e.g., [2, 3, 4]. Related issues have been studied previously in a variety of fields including organizational design [5], distributed and group decision making [6, 7], human-automation interaction [8] and manufacturing systems [9]. Here, we focus on a quantitative analysis of the stability and overall performance of the decision-making aspects of an organization from a systems-theoretic perspective. We model the organization as a dynamic system of multiple decision-making models.

The representation of human decision-making is a crucial aspect of our intellective models of decision-making organizations. For our purposes, we require highly abstract representations that reasonably approximate human behavior without reference to the semantic content of particular decision-making tasks. The representation we choose is based on the fact that decisions (and all other cognitive processes) take time. Thus, decision quality (e.g., accuracy) decreases when decision-making environments dictate that decisions be made before decision processes can be fully executed. More specifically, the accuracy of human decision-making for a particular decision-making task decreases in a nonlinear fashion as the rate at which decisions must be made increases. Demonstrations of this trade-off between decision speed and decision accuracy are widespread in experimental psychology. Furthermore, the behavioral research base overwhelmingly shows this trade-off to be S-shaped [10, 11], as shown in the notional plot of decision accuracy as function of decision workload (i.e., rate) in Figure 1. There is nothing magical about this S-shaped function. It merely shows a soft threshold for the impact of time-pressure on decision quality (i.e., accuracy). The negative acceleration at the tails of curve simply indicates that the effect of time-pressure on decrements or increments in decision quality diminishes as decision accuracy approaches the limits of 0 and 1. It should be noted, however, that conclusive evidence of speed-accuracy trade-offs in more complex decision tasks (e.g., team decision-making) is scarce. Given the methodological difficulties associated with demonstrating speed-accuracy trade-offs in basic laboratory tasks, the lack of evidence in complex tasks is not surprising, but one may conjecture that the S-shape tradeoff also applies.

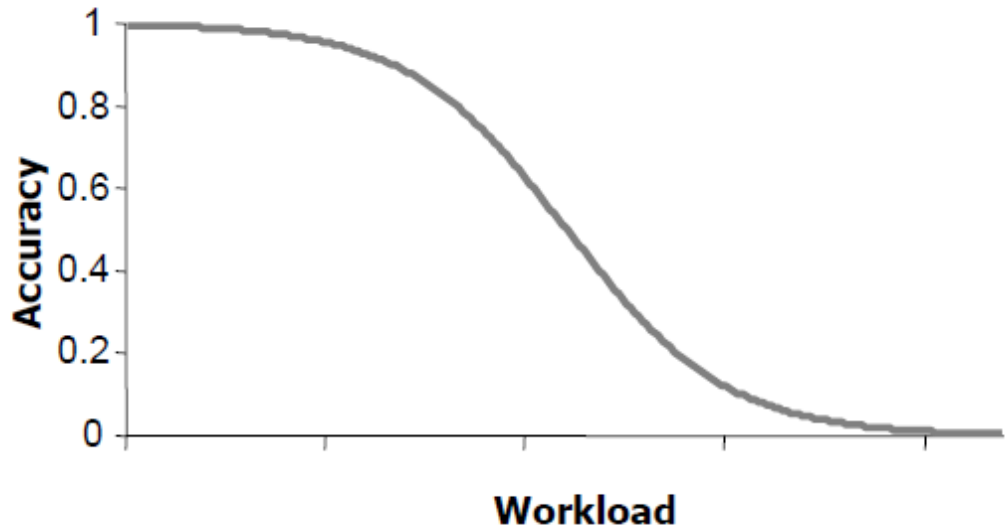

**Figure 1. Accuracy-workload trade-off curve for human decision-makers is commonly described by an S-curve.**

The rate of requests for decisions is only one measure of decision workload. Other factors contributing to greater workload may include, for example, the complexity of decisions; criticality or risk associated with decisions; uncertainty in the available data; and latency of the available data. Arguably the impact of these factors should be accounted for in models of decision-making organizations. However, the veracity of such arguments must be considered with respect to the goals of the modeling endeavor. Decisions in dynamic environments do indeed engage a variety of interrelated cognitive processes, ranging from monitoring, recognition and information search to planning, judgment and choice. Incorporating these processes into a simulation to account for the variety of factors that may influence decision workload requires a commitment to a particular model of decision making processes – a topic of continuing debate. The indisputable fact that remains is that each of the cognitive processes engaged by dynamic decision-making tasks takes time. Thus, in dynamic environments decision-makers are placed in a situation where they must control one time-dependent process (i.e., the evolving business situation) with another time-dependent process (i.e., the cognitive processes underlying dynamic decision-making). Decision-makers face this situation regardless of the complexity of a decision and any risks associated with it, or the quality of the data on which that decision is based. For our purposes, therefore, we focused on the simplest and arguably the least contentious measure of decision workload – the rate of the decision requests per unit time.

Consider a model of a decision-making organization consisting of a *Head-Quarters (HQ)* component and a *field* component as illustrated in Figure 2. The HQ component receives an input flow of orders from a higher authority ($u$ – the number of orders per unit time) as well as a flow of requests for decisions from the field component ($x2$). The HQ component produces a flow of commands ($x1$) and sends them to the field. In general, some of the commands may be erroneous. A workload-accuracy trade-off function $f(x)$—an example of the S-curve discussed earlier—governs the fraction of the errors. If a command is correct, it is assumed that the field component executes it successfully. If the command is erroneous, it results in problems in the field. The problems manifest themselves in the number of requests for decisions generated by the field component and sent back to the HQ component. A constant coefficient, $K$, relates the number of erroneous commands to the number of new decisions that must be made as results of the errors. A greater value of $K$ corresponds to a greater confusion caused by an erroneous command within the field component, and to a greater ability of the adversary to exploit the error. Here, a discrete-time approach is used; the accuracy, at a particular time instant, of outgoing decisions is a function of the decision requirements at the previous time instant.

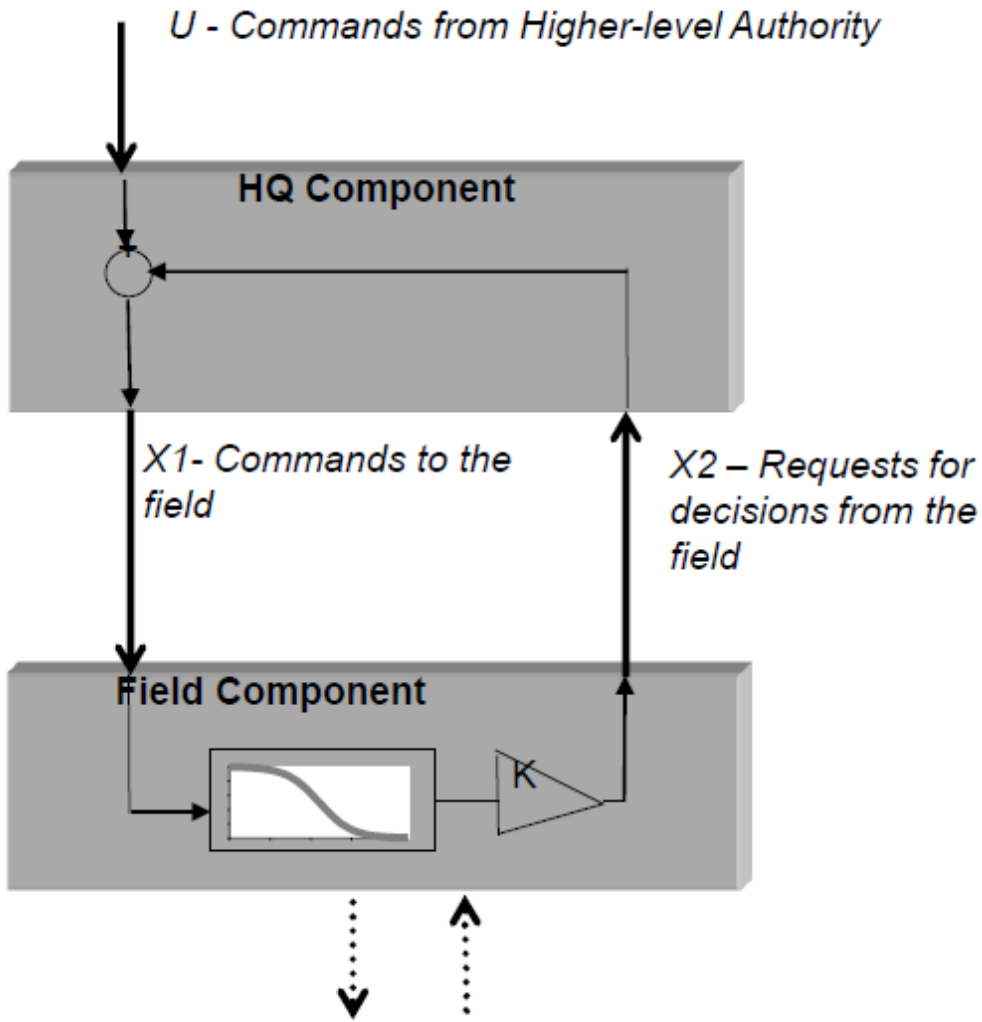

**Figure 2. A decision-making organization consisting of a Head-Quarters (HQ) component and a field component.**

If we use *x1* and *x2* as internal states, the dynamics of the system are given by

$$x_1(k+1) = x_2(k) + u(k)$$

$$x_2(k+1) = x_1(k) \cdot K \cdot f(x_1(k))$$

where

$$f(x) = 1 - \frac{1}{1 + e^{(x-a)/b}}$$

Linearization yields a sufficient condition for stability of the original nonlinear system:

$$K < 1 \text{ or } -1 < [1 - \frac{(1 - x_1/b) \cdot e^{(x_1-a)/b} + 1}{(1 + e^{(x_1-a)/b})^2}] \cdot K < 1, \tag{1}$$

and equilibrium points $\bar{x}_1, \bar{x}_2, \bar{u}$ must satisfy

$$\bar{u} = \bar{x}_1 \cdot (1 - K \cdot f(\bar{x}_1)) \tag{2}$$

$$\bar{x}_2 = \bar{x}_1 \cdot K \cdot f(\bar{x}_1).$$

Numerical computations using Matlab [12] yield the results depicted in Figure 3. The lines marked "20%" and "80%" show where the fraction of the erroneous commands issued by the HQ component stays at the levels of .2 and .8 respectively. We observe that for *K*<1 the system remains theoretically stable, but higher values of *u* can lead to a rapid increase in the fraction of the erroneous decisions, in essence supplying a domain-specific type of instability. For *K* >1, the system can exhibit unstable behavior at higher values of *u*. As *K* increases, the instability occurs at progressively lower values of *u*. In domain-specific terms, this means that if one lowers the ability of the field component to correct for erroneous commands, then the combined decision-making organization becomes unstable at lower values of input commands (*u*).

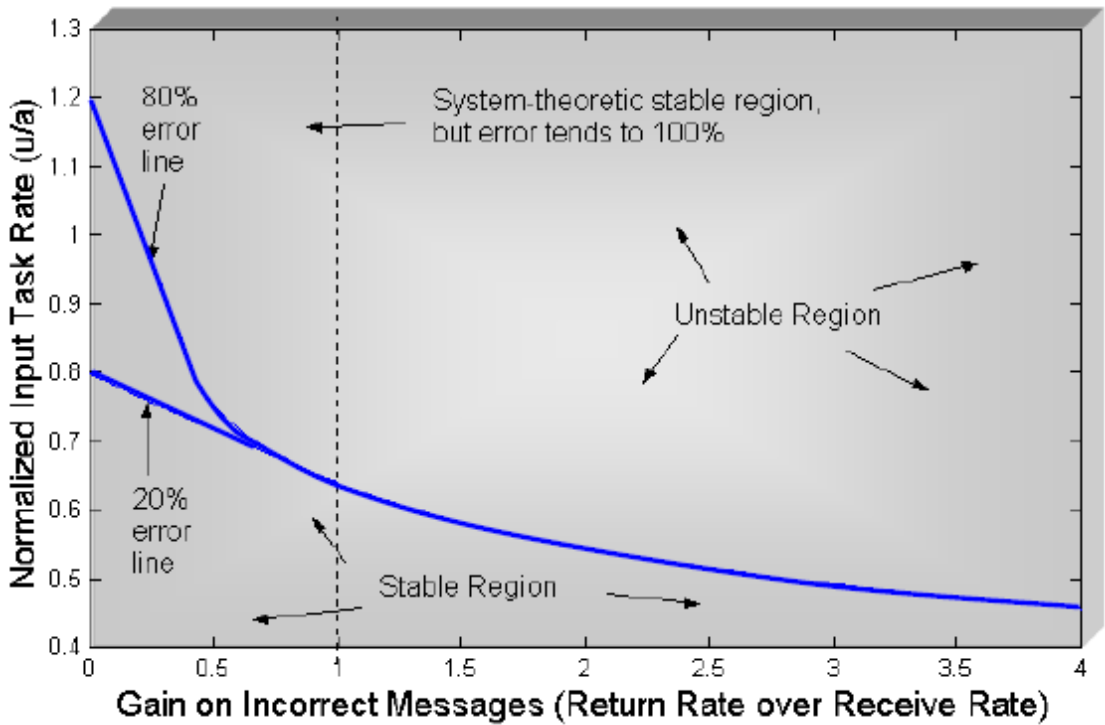

**Figure 3. Stability regions as a function of gain on incorrect messages and of input task rate.**

Clearly, this result is qualitatively consistent with the realistic scenarios A and B we introduced above. The value of the model is not in its ability to predict specific quantitative behavioral characteristics of a specific organization. Rather, it offers easily comprehended suggestions of possible phenomena to explore with more in-depth analyses (e.g., with high-fidelity emulative models). So, what does this model suggest?

First, it points to the importance of modeling feedback channels, particularly those associated with erroneous decisions. Depending on the structure and tasks of the organization, the impact on its productivity can vary widely.

Second, our model highlights the importance of modeling the reduction in decision quality due to time pressure. The drop in quality of decision making due to the complexity of a task can have a massive impact on the predicted performance of an organization.

Third, our model allows us to hypothesize useful directions for further study in terms of the design of experiments that employ high-fidelity emulative models. In particular, the susceptibility of an organization to self-reinforcing error, according to this simple intellective model, depends mainly on two key parameters: $a$ –the extent to which the decision-maker is able to absorb the flow of higher-authority commands, and $K$—the measure of how well the field operators are enabled and empowered to handle locally erroneous or late decisions coming from their superiors. Knowing these major influencers can be valuable guidance for a designer of an organization who models its performance in order to verify its robustness under dynamic conditions.

*Propagation of Disruptions in Organizations*

Now let us extend the simple model of the previous section to consider networks of decision making units. In order to do this, we first build up a generic component model that can be inserted in a standard hierarchical authority structure where tasks or messages are received from a superior and distributed amongst subordinates. Such a model is illustrated in Figure 4.

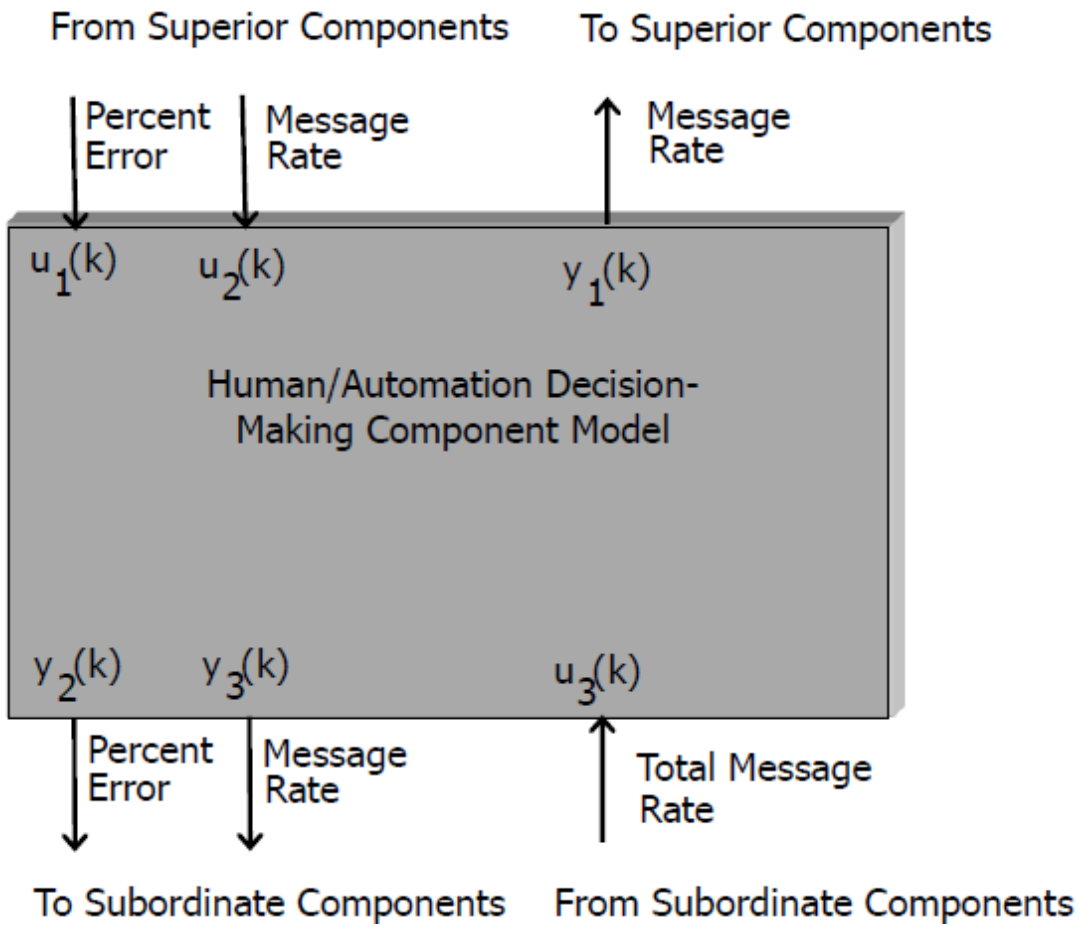

**Figure 4. An intermediate component in a network of decision makers.**

In this model, message flow down through the hierarchy is characterized by the rate and accuracy (percent error) while message flow up is characterized only by the message rate. We emphasize that this model is also intellective, as in the first example above. The incoming message load to this component is $u_2(k) + u_3(k)$, and we assume that the errors created by this component are based on the same speed-accuracy trade off in the first example above, i.e. the incurred error percentage is

$$f(u_2(k) + u_3(k)), \quad \text{where} \quad f(x) = 1 - \frac{1}{1+e^{(x-a)/b}}$$

This error percentage is in addition to that already in the messages from the superior, $u_1(k)$. The rate of messages to the superior is meant to capture the need for clarification, and therefore we assume this is the product of the current incoming rate from the superior, the total error percentage of those messages $u_1(k) + f(u_2(k) + u_3(k))$, and a gain factor K representing the susceptibility to confusion at this component. We assume the percent error in the messages that flow through to subordinate nodes is the sum of the received error rate from superior nodes, $u_1(k)$, and the incurred errors by this component in the correct messages, $(1 - u_1(k))f(u_2(k) + u_3(k))$. And, finally, the message rate to the subordinate nodes is set equal to the received message rate from the superior node at the previous time step. Note that the message rate received from subordinates, $u_3(k)$, may also include exogenous inputs representing additional requests for decision directly from the field.

This results in the following set of difference equations.

$$y_1(k+1) = K[u_1(k) + f(u_2(k) + u_3(k))]u_2(k),$$

$$y_2(k+1) = u_1(k) + (1 - u_1(k))f(u_2(k) + u_3(k)),$$

$$y_3(k+1) = u_2(k),$$

where again

$$f(x) = 1 - \frac{1}{1+e^{(x-a)/b}}$$

This model of an intermediate component is used to construct networks of decision-makers by linking inputs and outputs. We consider first the network illustrated in Figure 5. This is a purely hierarchical network in which components receive and interpret messages from their superior components and then direct messages to their subordinate components (as well as respond to their superiors) as expected from the description above.

The stability of the hierarchical network was analyzed in simulation, subject to the rate of exogenous high-level inputs from above, i.e. $u_1(k)$ in the top component, and the rate of exogenous inputs from the field, i.e. $u_3(k)$ in the four bottom components. Figure 8 shows the average error rate in commands to the field components, i.e. the average of $y_2(k)$ in each bottom component, as the exogenous input from above is increased. We note that the system maintains its performance at a near-constant level and then rapidly collapses.

To evaluate the envelope for stability subject to variations in exogenous inputs from the field, we stimulated the inputs $u_3(k)$ in the far right and far left components with constant inputs. Figure 9 shows the stability envelope subject to the sum and difference of these constant inputs. The interpretation of *stability* in these graphs is not a standard systems-theoretic instability (i.e. bounded inputs result in bounded outputs) but rather a domain-specific meaning of instability alluded to above (i.e. feedback leading to the production of high fraction of incorrect commands). In these simulations, systems-theoretic instability was excluded with the use of saturation devices. It was also observed during simulations of this network that the instability had a tendency to propagate through the hierarchical structure, i.e. overload at one location affects other locations by overloading the higher-level components which then in turn overload their subordinates.

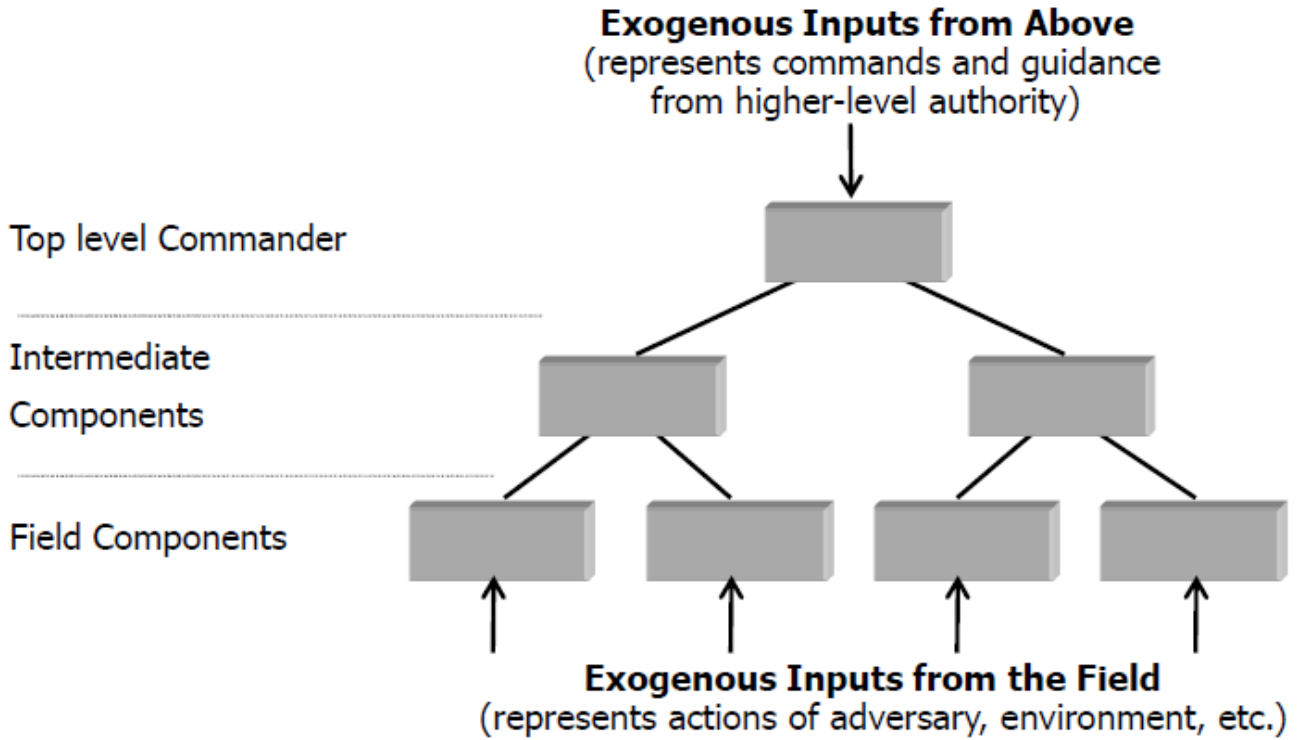

**Figure 5. A hierarchical organization.**

*Active Compensation*

Monitoring and diagnostic functions—human or computerized—are critical to detection and mitigation of positive feedback. The intent of such monitoring is to introduce corrective measures when the system or organization approaches a danger zone, perhaps the limit of stability. We also mentioned the possibility of using reconfigurable systems to deal with undesirable and non-normative situations. Such ideas lead us to consider a possible approach to mitigating degradations in decision quality based on an active compensating component.

Consider the network illustrated in Figure 6, in which a compensating component, e.g. a task allocator or broker that re-distributes messages or tasks, is introduced to the network as an intermediary between superior and subordinates in order to re-distribute tasks or messages. The compensating component is illustrated in Figure 7. Although this component continues to propagate errors through the system, it attempts to drive the message rates to subordinates to a uniform distribution of messages, i.e. based on measured return rates from the subordinate components, future messages are directed away from components that have high-levels of additional decision-making requests. It is assumed that the compensating component is itself subjected to the speed-accuracy trade-off because any functions that were performed by the intermediate levels in Figure 5 remain. Hence, the redistributing component adds errors to the messages depending on the total number of messages received from superior and subordinate components. This also allows us to focus on the impact of the redistribution of messages rather than the impact of the removal of the intermediate level of decision-making. The following difference equations are used to model the behaviour of the compensating re-redistributing component. .

$$y_1(k+1) = u_1(k)[f(u_1(k) + \sum_{i=3}^{n} u_i(k))(1 - u_2(k)) + u_2(k)]$$

$$y_2(k+1) = f(u_1(k) + \sum_{i=3}^{n} u_i(k))(1 - u_2(k)) + u_2(k)$$

$$y_i(k+1) = u_1(k)(\frac{n-4}{n-3})(1 - \frac{u_i(k)}{\sum_{j=3}^{n} u_j(k)}), \quad 3 \leq i \leq n.$$

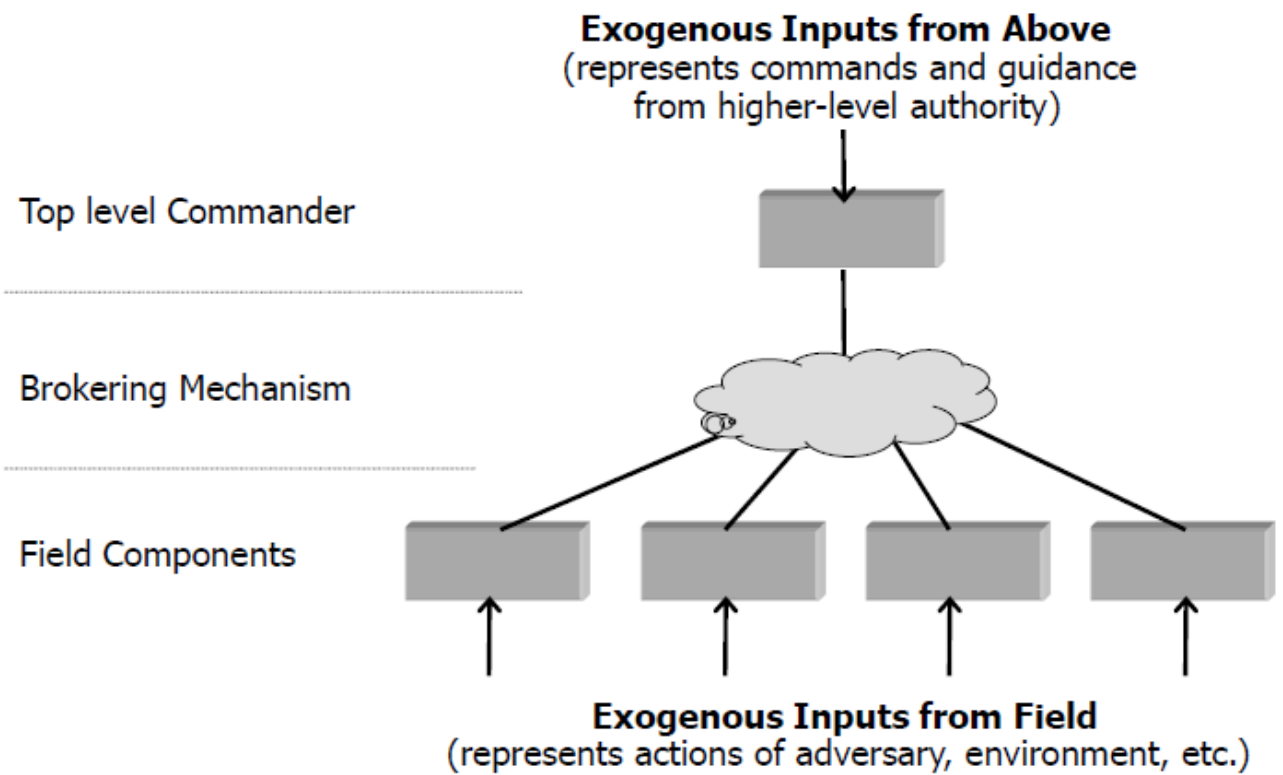

**Figure 6. A network with a compensating component (e.g. a broker).**

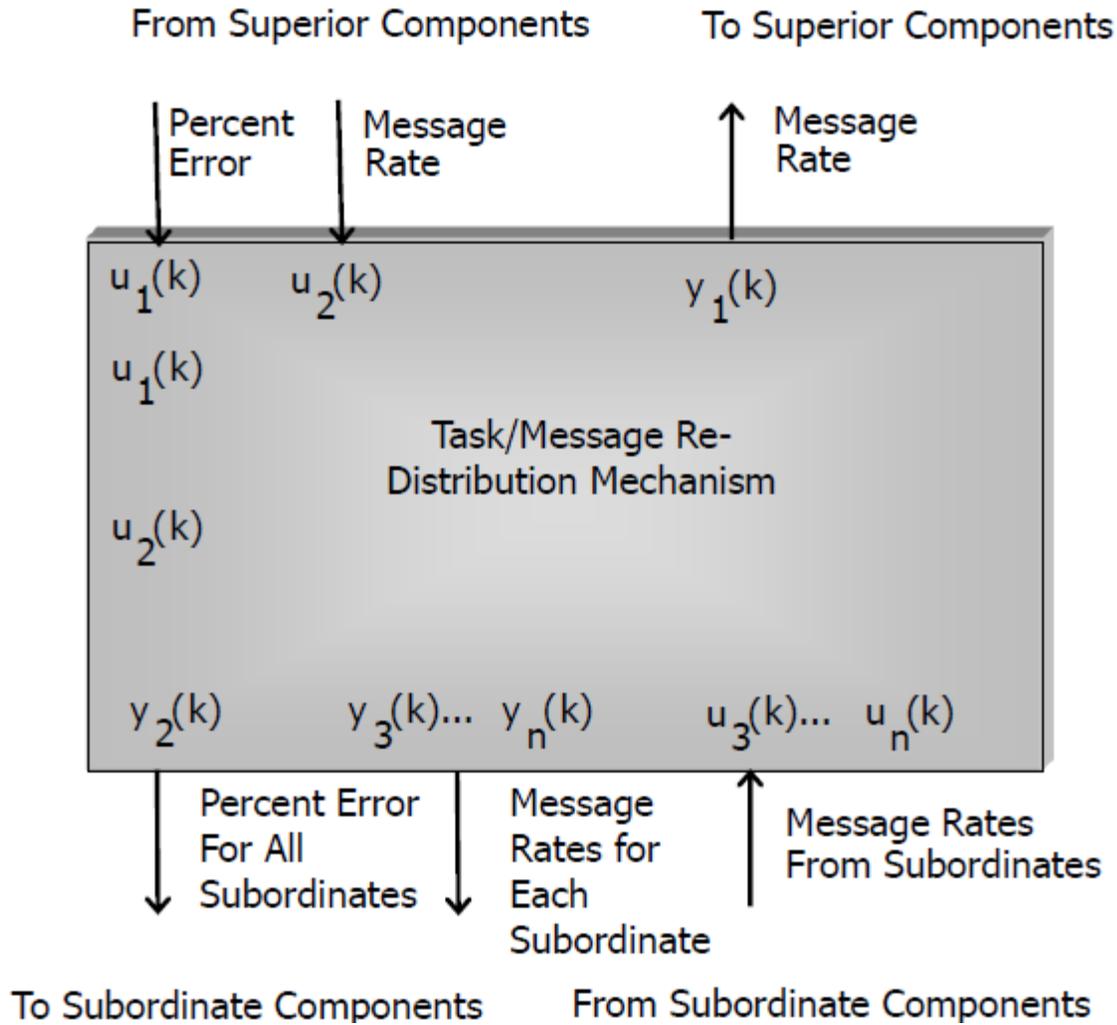

**Figure 7. A compensating component.**

Again, the stability of the network with a redistribution component was analyzed in simulation subject to the rate of exogenous high-level inputs from above, i.e. $u_1(k)$ in the top component, and the rate of exogenous inputs from the field, i.e. $u_3(k)$ in the four bottom components. Figure 8 shows the results along with the results for the hierarchical network. Again, the graphs shows the average error rate in commands to the field components, i.e. the average of $y_2(k)$ in each bottom component, as the exogenous input from above is increased. We note that with the redistribution component, the system also maintains its performance at a near-constant level and then rapidly collapses, though now at a higher total rate than the pure hierarchical network. Again, the stability was evaluated subject to variations in exogenous inputs from the field in the same way as for the hierarchical network, i.e. inputs $u_3(k)$ in the far right and far left components were stimulated with different constant inputs. Figure 9 shows the stability envelope subject to the sum and difference of these constant inputs, alongside that for the hierarchical network. The stability envelope is increased with the inclusion of the compensating component, and therefore the system becomes more robust to variations in the difference in exogenous inputs at different components.

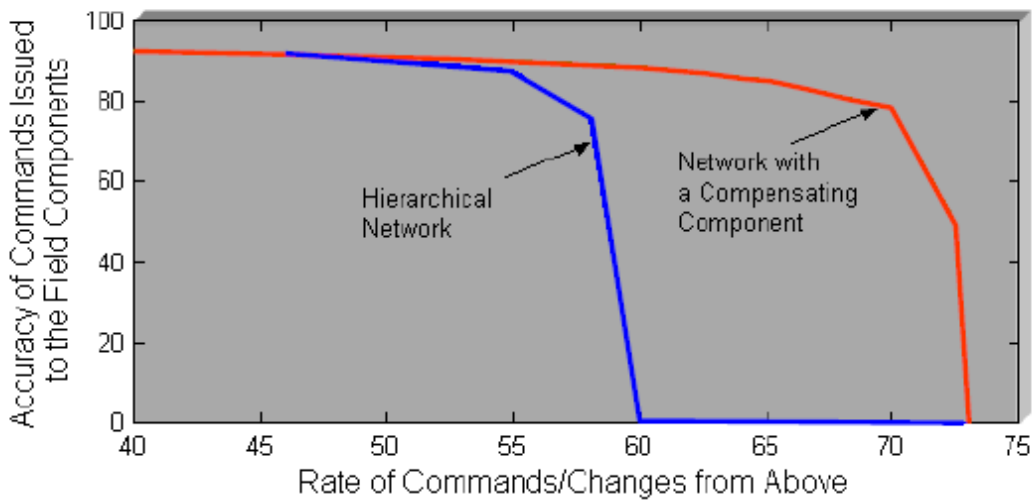

**Figure 8. Accuracy of messages to field components as a function of the rate of high-level commands.**

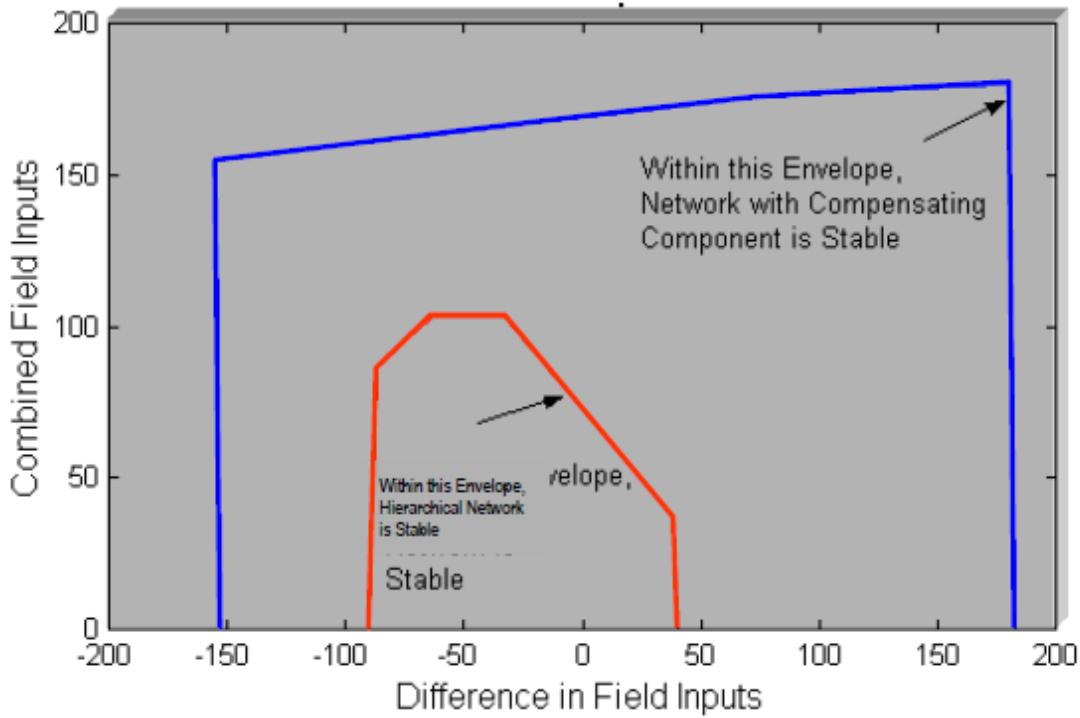

**Figure 9. Stability regions with respect to combined field inputs and difference in field inputs at two locations.**

*Dynamic Reorganization to Mitigate Malfunctions*

The simplified models above indicate that an appropriately designed compensating component could significantly improve the performance of a decision-making organization over a broader range of operating conditions. But so far we said nothing about how such a component would actually be implemented. We explore now a more specific scheme to such an active compensation, and consider the internal workings of a compensating component that dynamically re-organizes the allocation of decision-making responsibilities and flows of relevant information.

In the traditional theory and practice of distributed and hierarchical control systems, a supervising controller issues commands to lower level controllers. These commands become inputs or set points for the control processes of the lower-level controllers. The responsibilities and types of interactions between lower level controllers are static. They are fixed when the system is designed and remain constant while the system operates. When this constant definition of responsibilities and interactions becomes inadequate to adapt to the changing circumstances, as will be the case in an uncertain environment with adversarial threats and changing information requirements, the control system is unable to perform effectively. Additionally, in the context of a team of decision-makers that is composed of humans as well as automated control processes, it is typically a human senior leader (a military commander, or a senior manager) who controls the allocation of decision-making and information-sharing responsibilities as well as the modes of such sharing. The structure of the team is also often fixed in a hierarchical format. A human leader typically adjusts decision-making and information-sharing responsibilities based on informal techniques, intuition, and best guesses based on prior experience and training [2, 5, 13, 14]. However, this control can break down when the complexity and speed of changes in the situation exceeds the cognitive and reasoning capabilities of the human controller. This problem may be exacerbated in those teams that include software agents and robots as decision-makers [E.g., [24]) because these artificial decision-makers can observe, execute their decision-making algorithms, and act much faster than a human controller, potentially leading to a cascade of failures and poor team performance [15].

Here we explore a computer-assisted approach for managing decision-making, information-sharing responsibilities, and modes of interactions between the team members that allows the team to effectively and rapidly adapt to changing circumstances, threats and opportunities.

The predictive control scheme is illustrated in Figure 10 and involves four main aspects that are described in the following sections. Predictive control is often employed, even if only implicitly, in planning and execution of military operations and industrial engineering applications. An internal model exists, even if only informally in the commander's situational awareness, and is used to determine an input that can then be applied to the real system. This is done either analytically if it is a mental model or through simulation if it is a computer model.

In Figure 10, we consider the input to the physical world or action space, to be the *structure* of the decision-making team, i.e. the decision responsibilities and information channels. The epistemological content of the decisions themselves remains in the physical action space. The internal model within the predictive controller is only an abstraction that captures quantitatively the decision load and information channel loads. Similarly, as illustrated in Figure 10, the forecast decision requirements from the physical action space are abstracted to a set of time-varying parameters that are used as an open-loop input for the simulation of the decision-making process. The standard model-predictive control process is then enacted; different information structures are simulated with the given forecast decision requirements and a best structure is chosen and used for the real decision making team in the physical action space. This process of simulating a variety of information structures is captured by the internal optimization loop of Figure 10.

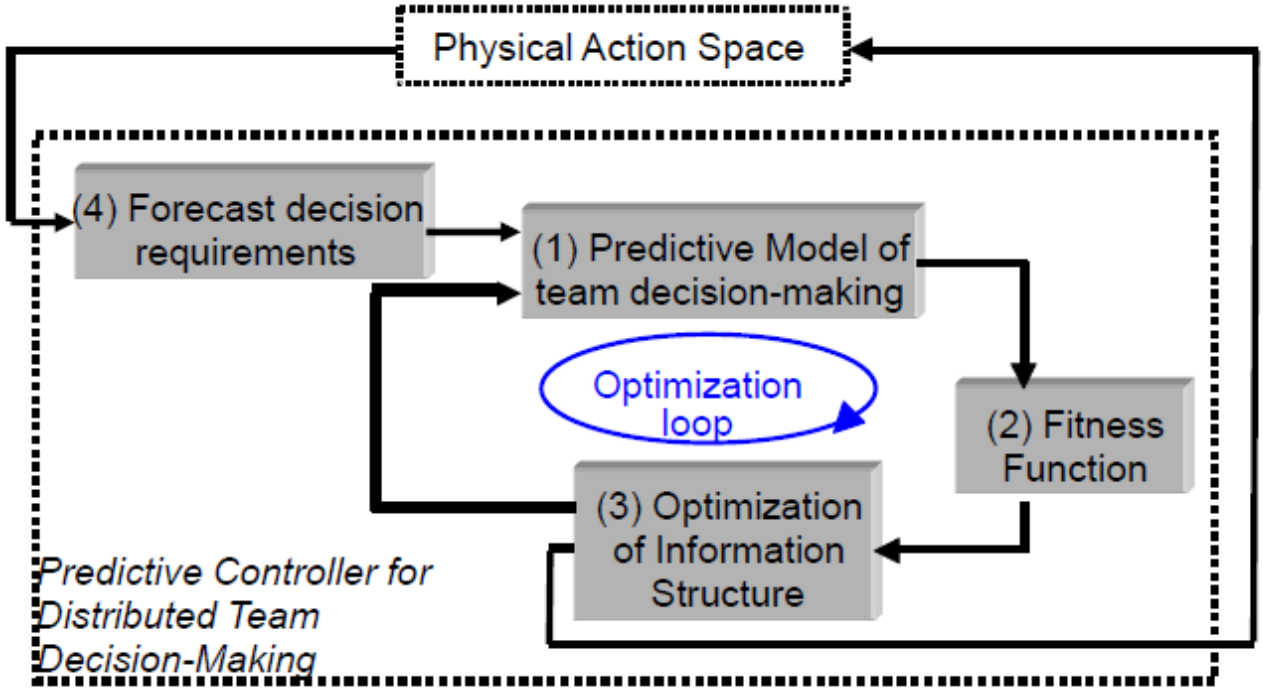

**Figure 10.** The control scheme for dynamic allocation of decision responsibilities and information sharing.

*Modeling Team Decision-Making: Decision Responsibility and Information Structure*

For illustrative purposes, we consider a simple example of a team of three decision-makers – a foreman, a scout and a robot – involved in a mission whose objective is to extinguish a fire in a large chemical plant. We assume that an initial ad-hoc plan calls for the foreman to observe the fire from an observation post, the scout to identify a target location for dropping fire extinguishing material and then join the foreman, and the robot to drop the fire extinguishing material at the identified target location. An example of possible decisions required is as follows:

U1 – When and where to call a vehicle for the foreman escape.

U2 – When and on which target to drop fire extinguishing material.

U3 – Which flight path to take to reach the target.

U4 – When to egress and which egress route to choose for the scout.

We will re-visit this example as notions are developed. The questions we wish to address in the context of this example are

- Which, of the three team-members, should be responsible for deciding U1 through U4?
- How should the current decision outcomes be disseminated to other members in the team (noting, of course, the key fact that too much information may have negative consequences when an urgent decision is needed)?
- If we can predict the difficulty and urgency of decisions U1 through U4, can we manage the answers to the first two questions? i.e. can we dynamically update who is responsible for which decisions and how the decision outcomes are disseminated.

Note that the decision-makers in this example are one and the same as the actors, but this is for convenience only. There may be entities in the decision-making teams that are physically removed from the field, e.g. an experienced foreman at another plant. Also note that here we are not considering the control of the observation process, again for convenience. Elsewhere [23], it is proposed that observation channels be controlled in much the same way as decisions. In our example, this could mean for instance appending the observations 'Z1- Location of the fire' and 'Z2 – Types of target locations' to the list of decisions.

Let us employ a straightforward model to capture the traditional relationship between the task characteristics of complexity, urgency and decision load and the resulting accuracy of the decision [10, 11, 16, 17]. A set of parameterized curves in which accuracy is inversely proportional to the time pressure is used for this purpose. An example of this relationship is shown in Figure 11. Unlike the simpler S-curve (Figure 1), here the additional parameters of normalized discriminability and the number of the options provide an upper and lower bound on the accuracy as illustrated. In a practical implementation of the scheme, characteristics of the decision-making entities could be used to quickly generate models from a pre-determined virtual decision-maker coefficient database. Such characteristics might include rank and experience level for a human decision maker or processor capability, function and speed for an artificial decision-maker. As we are presently interested in the analysis of trends and the effectiveness of the approach, a relatively simple model suffices.

The relationship in Figure 11 is used to build dynamic input/output models for single decision-makers in the same way as was done in Figure 4 for the previous example The normalized discriminability and number of options are exogenous inputs to the decision-making team and we assume these are properties of the decisions themselves rather than the interaction processes in the team decision-making. Further, we assume the time-pressure has two sources. First, there is an exogenous component that must be predicted, but additionally there is time-pressure that results from the necessity of interactions with other team members. Clearly, if another team-member is urgently soliciting information on a different topic, this reduces the amount of time that can be spent on the decision at hand and so increases the time pressure.

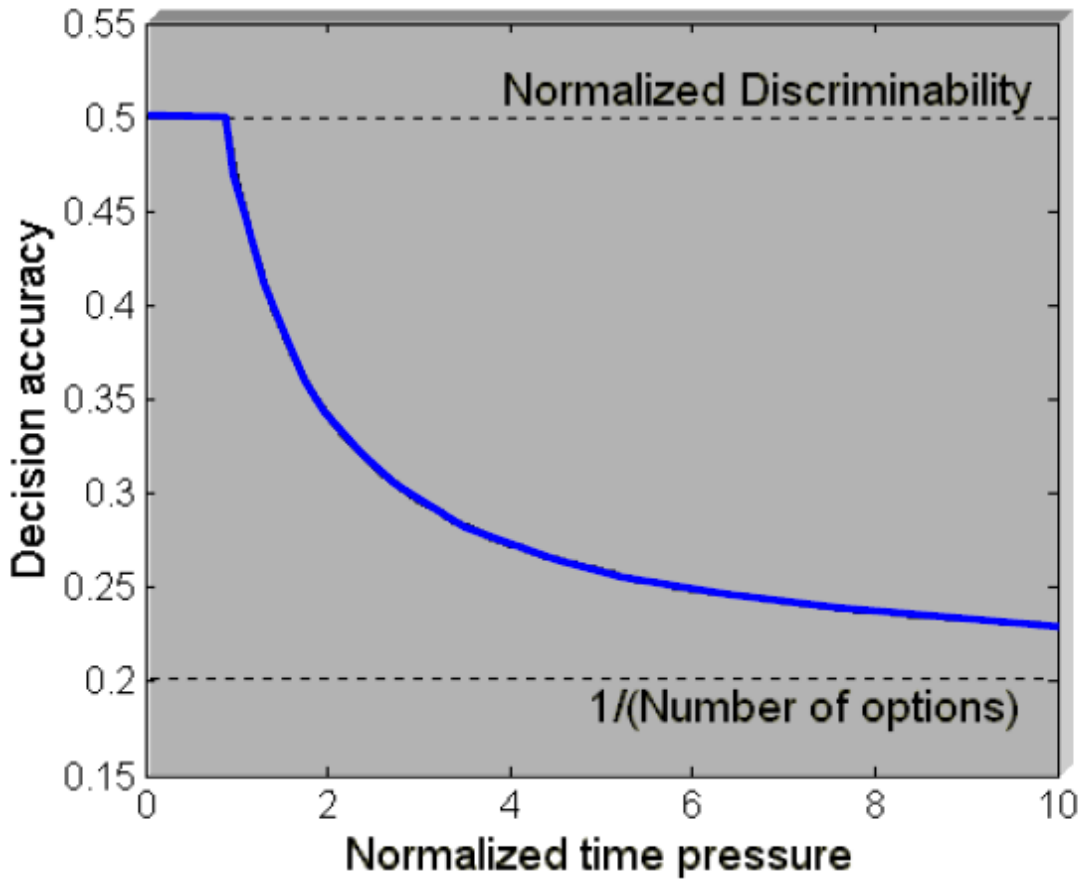

**Figure 11. The decision-accuracy relationship for one decision-maker.**

The input/output model for a single decision-maker is used as an atomic block to construct the team decision-making model. The connections between the atomic blocks are determined by what we will term the *information structure*. This term is not uncommon in the literature on decentralized and distributed control in more general settings [18, 19, 20]. We borrow the term here to refer specifically to the combination of three objects: an *observation structure*, which maps the incoming data streams to the decision-making entities, or agents, a *decision-responsibility structure* which partitions the outgoing decision variables amongst the agents and a *decision-sharing structure* which defines the information paths between agents.

A possible information structure for the fire-fighting example discussed above is partially illustrated in Figure 12. The decisions that each member is responsible for are indicated beside each agent, e.g. the foreman is responsible for U1, and this defines the *decision-responsibility structure*. The *decision-sharing structure* is illustrated with connecting arcs. The labels on these arcs show the decision that is being communicated and the *mode* in which it is communicated. Here, we consider only two modes: a *pull* mode in which the first decision-maker communicates information only if the second requests the communication, and a *push* mode, in which the first decision-maker immediately communicates any new information. Though there is certainly a spectrum of possible interactions of this sort, we have identified these two modes for a preliminary analysis. The observation structure is again left out for convenience (see the previous footnote). Note, from a semantic point of view, there is a minimal necessary information transfer for the decisions to be made. For instance, decision U3 (the route taken to the target location) necessitates knowledge of decision U2 (selection of the location and timing for the target). Accordingly, we define a *decision dependency criterion* that underlies the dynamics of the model. Information structures that do not meet this minimal criterion will perform poorly because, for example, decisions concerning U3 made in the absence of knowledge of U2 are given arbitrarily an accuracy of zero per cent (we assume informally that two wrongs do not make a right). The decision dependencies for this example are that U1 requires knowledge of U2 and U3, U2 requires U1, U3 requires U1 and U2 and U4 requires U1 and U2. We assume that the minimal criterion can be met with either the pull or push mode communication. For simplicity, the observation process is not modeled in this example, and so the *observation structure* is not defined here. However, in general, the observation structure would define a set of observations, e.g. Z1 and Z2 for each decision-maker.

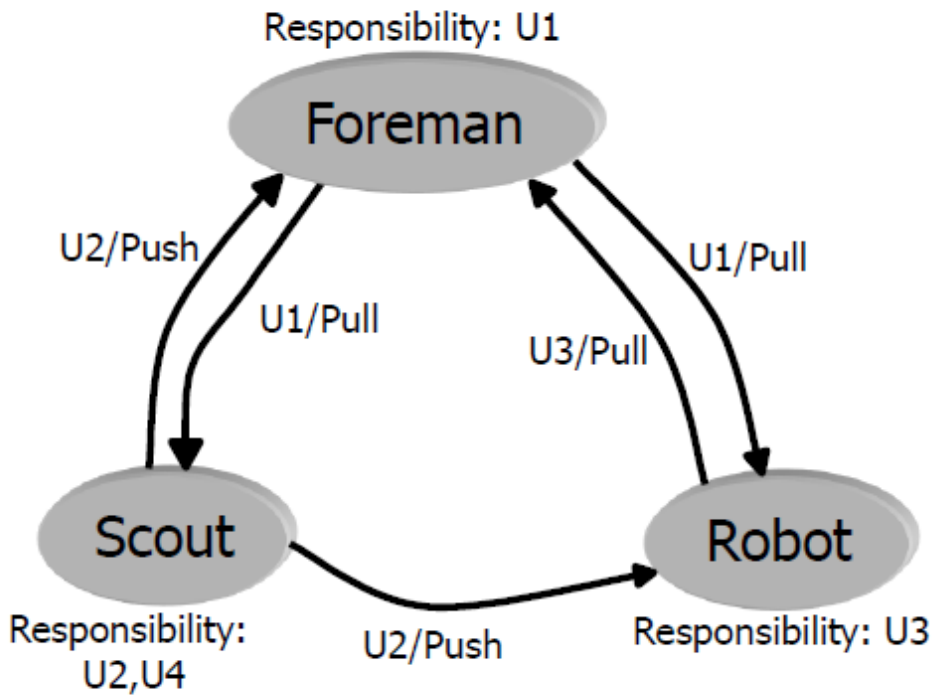

**Figure 12. The decision-makers in the fire-fighting example communicate with one another and take responsibility for specific tasks as described by an information structure.**

*Measuring Decision-Making Performance*

How do we know if a one information structure is better than another? A quantitative measure of decision-making performance (e.g. overall accuracy or timeliness of decisions) is required as a measuring stick for the suitability of a given information structure. Clearly, the true intrinsic value of a given information structure is its effectiveness in the real world, but unfortunately without a high-fidelity simulation of the "physical action space" in Figure 10, this is not available. While a number of surrogate measures can be envisioned, we explore a particular one - a weighted average of the accuracies of all decisions being made by the team, integrated over the time horizon. This serves as an approximate indicator of performance. A potential implementation would couple decision-making models with an action space simulation so that the predicted overall effectiveness could in fact be used as a fitness function.

As indicated in Figure 10, the value of the fitness function is used to fine-tune the information structure. Smoothness or monotonicity of the fitness function with respect to changes in information structure will not generally hold. More importantly, changes in the information structure are discrete yielding a discrete (in state) and highly non-linear optimization problem. In general, traditional continuous-state feedback control techniques are not directly applicable.

Repeated simulation for different alternative information structures can be performed to maximize the fitness function. This in effect requires a search of the space of information structures. The optimization problem can be stated formally in a form similar to the observation problem suggested in [20] and here, as there, it will generally bear no analytic solution. In practice, the optimization of responsibility/information structure cannot be de-coupled from the optimization of the physical actions, such as the execution of a military operation. However, an iterative or serial approach may provide reasonable solutions; current traditional operations planning tools do not explicitly consider the information structure as a quantity to be controlled, however future operations planning may plan both physical operations and the information structure.

*Forecasting Decision Requirements*

In the decision-making accuracy curve in Figure 11, normalized discriminability, the number of options and a time pressure characterize each decision. When this model is used in simulation within each dynamic model of decision-making, predictions for these characteristics are required in order to select the most appropriate information structure. The characteristics are predicted for a rolling time horizon (for the next 6 hours, for instance). Of course, these predictions should change due to action space events that are a result of the change in information structure. However, that is not considered here; the predictions are taken as fixed for the purposes of optimizing the information structure.

In practice, the raw information from which these characteristics could be predicted comes directly from the observations of the action space and other information and intelligence sources.

An example of predicted decision requirements for the four decisions for the fire-fighting scenario is shown in Figure 13 for a time horizon of 6 hours. For the purposes of our example, these were generated manually on the basis of the notional progression of the fire and the related actions of the fire-fighting team. For instance, the decision characteristics of U1 (corresponding to the decisions concerning the escape vehicle for the foreman) are such that there is little time pressure and many options at time zero, but as the fire progresses the time pressure increases (reflected by a decrease in the graph, which depicts time availability) and the number of options decreases steadily. This might be the case if escape vehicle options become scarcer as the fire approaches locations close to the foreman.

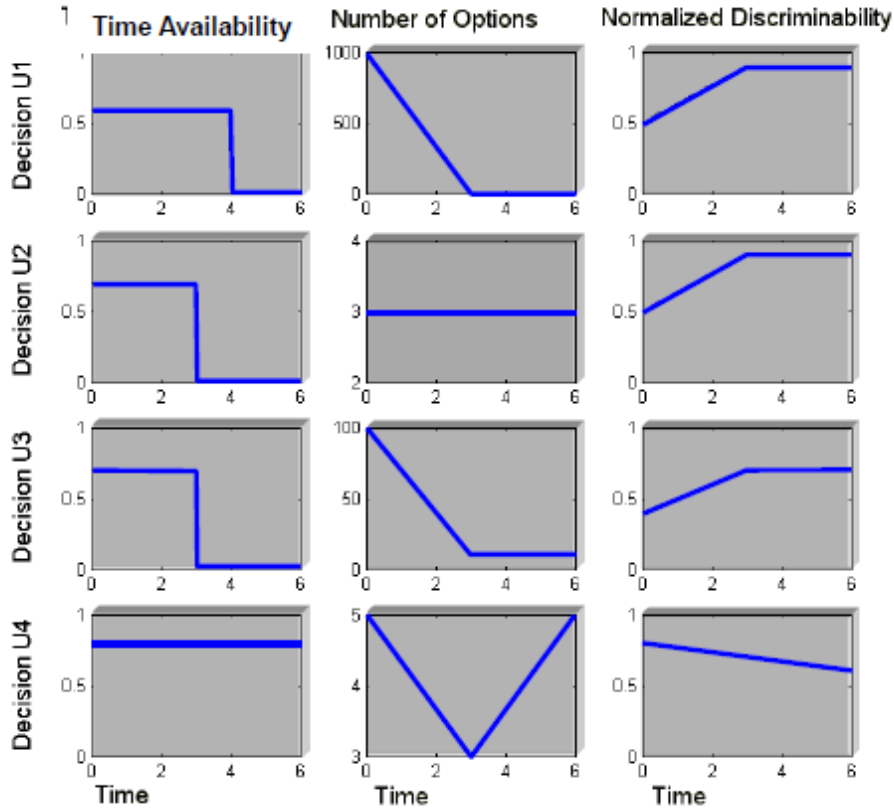

**Figure 13. Future decision characteristics are predicted from information sources and used to fine tune the information structure.**

*A Simulation of the Fire-fighting Example*

Figure 14 shows a Matlab Simulink [12] implementation of the model in which the atomic blocks for the foreman, robot and scout in the fire-fighting example are connected based on a given information structure, e.g. that in Figure 12.  The forecast decision requirements in Figure 13 are used as exogenous inputs to this model.  By simulating this model in time, we are able to generate, for any given information structure, an average accuracy of decisions over both time and decisions which provides a measure of performance for this information structure.

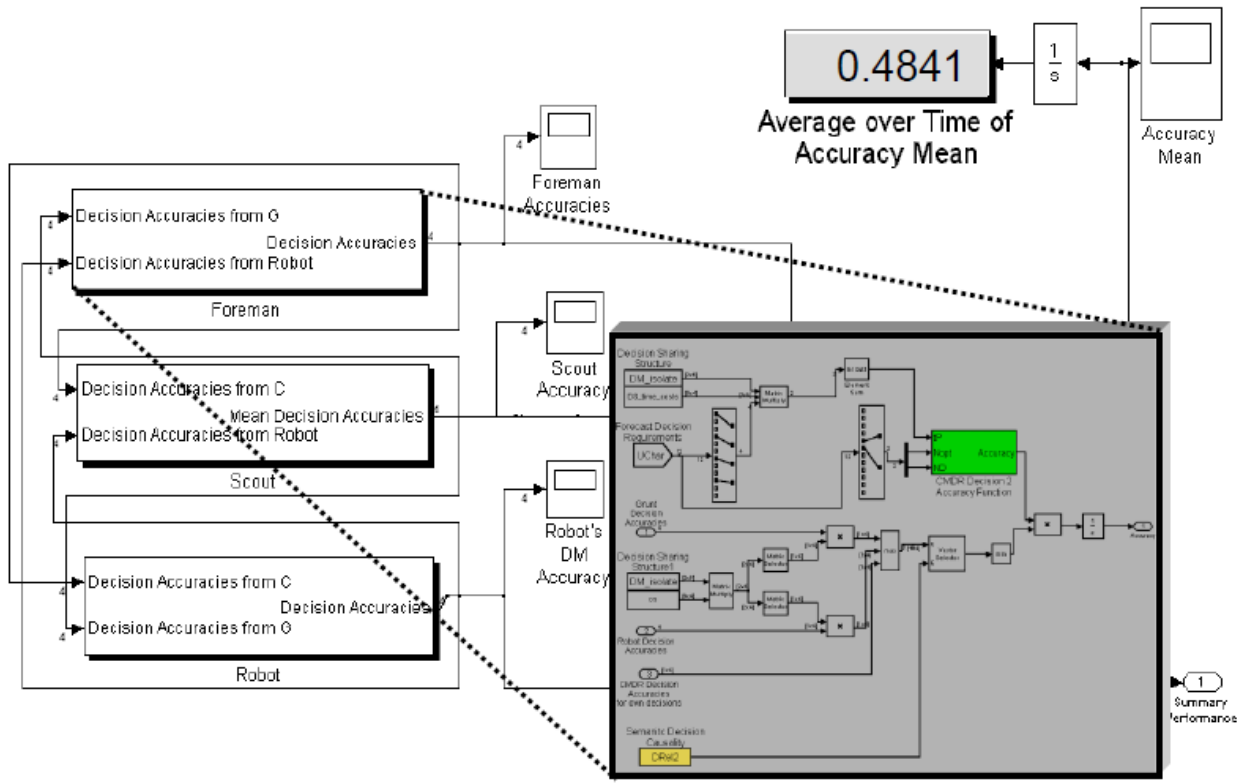

**Figure 14. The proposed scheme for predictive control implemented in MATLAB Simulink for a small-scale example.**

We investigated an optimization scheme based on a genetic algorithm, although a number of discrete optimization techniques could be potentially applicable. The information structure is encoded as a chromosome without content loss, and the fitness of each chromosome is evaluated with a direct simulation of the team-decision making model in MATLAB Simulink. The Genetic Algorithm Optimization Toolbox [21] offers a large variety of mutation operators, crossover operators and selection criteria and is well suited for this application.

Results for this optimization appear promising for the small-scale exploratory scenario with convergence to a (locally) optimal information structure with greatly improved fitness in realistic computational cost.

A small variety of meaningful, common-sense, information structures was produced (see Figure 15 for two examples), depending on how the forecast predictions were varied. This matches at least with intuitive notions of how teams should communicate and share tasks, and indicates the potential for the approach to be applied in real systems with higher-fidelity decision-making models.

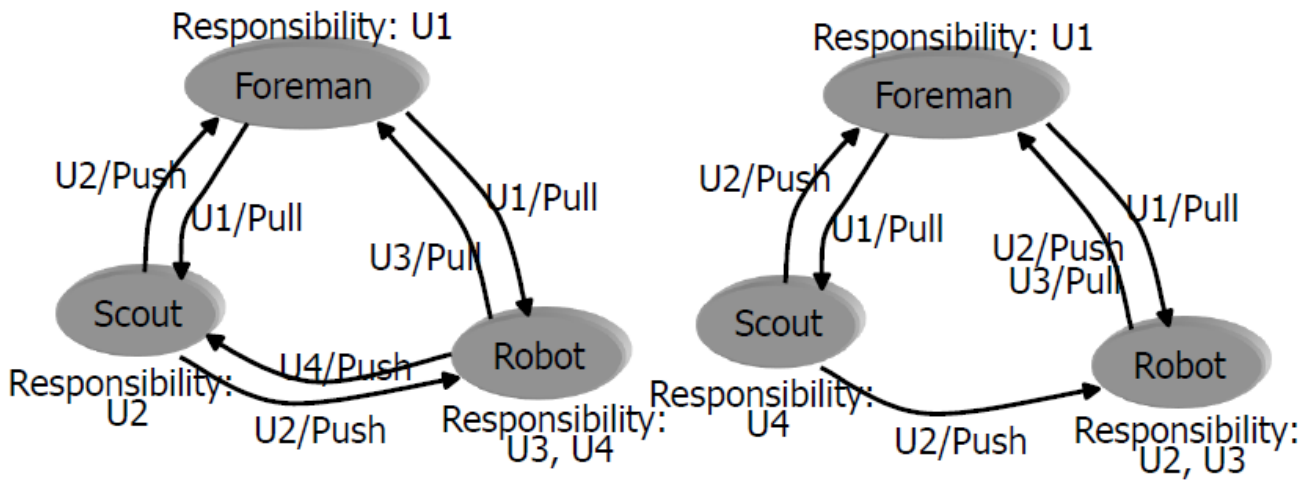

**Figure 15. Examples of locally optimal information structures for given sets of predicted decision characteristics.**

One might wonder how the effectiveness of such an approach depends on the characteristics of the environment in which the organization operates. Figure 16 conjectures a notional landscape of types of decision making environments categorized by the accuracy of the predictions that can be made and the rate of change of the environment or operations tempo. The regions denote where we believe it is appropriate to use dynamically managed team decision-making rather than traditional team decision-making where information structure is either rigid or evolves on an ad hoc basis. As the tempo of operations increases and the accuracy of predictions in the action space decrease, we believe it becomes more difficult to recognize the need for changes in information structure and therefore it is in these cases that dynamic management of the information structure will improve performance.

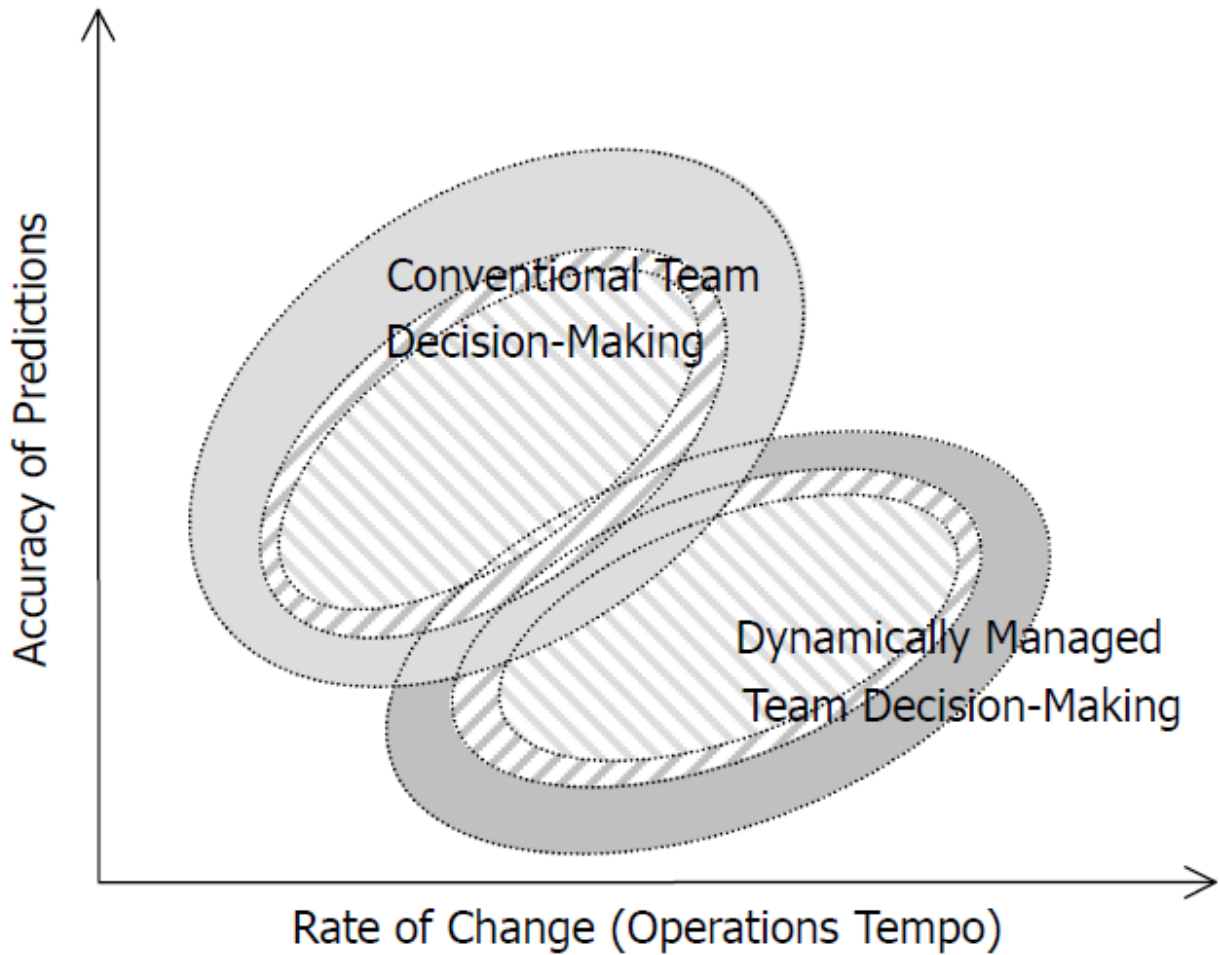

**Figure 16. The regions of general applicability of conventional and dynamically managed team decision-making.**

*Mitigating and Inducing Malfunctions*

So, what do these models and simulations tell us about the susceptibility of an organizational decision-making process to an adversarial action? What can an organization do to mitigate its susceptibility to a malfunction? What can an adversary do to induce a malfunction? Based on the results of the simulations we described, one can suggest several approaches to architecting and operating an organization in a way that reduces its susceptibility to malfunctions.

*Empower subordinates via mission-type orders.* The onset of collapse can be delayed by reducing the number of requests sent from the field components to the decision-making components This suggests that such a collapse can be mitigated by enabling the field components to operate as independently as possible in order to minimize the amount of cases in which they must call for the decision of the higher decision-making components. Providing the field components with maximum autonomy, using mission-type orders that specify the goals but not the specific ways to achieve them and using command-by-negation (defined below) all contribute to the reduction of the *K* parameter and strongly improve stability of the overall system (see Figure 3). Conversely, an adversary may exploit the situations where field components within the target organization are compelled to operate under strict and detailed control by higher-level decision making components. An adversary may also attempt to induce this tight control by generating extensive negative publicity about any error committed by field components.

*Prioritize and delegate.* Degradation can be avoided by dumping excessive messages, i.e. by ignoring some of them and insisting subordinate decision-makers handle problems autonomously. In practice, this is one of the mechanisms that organizations employ to reduce the effects of decision overload. Decision-makers learn to prioritize their decision-making load ignore what appears less important.

*Use command-by-negation, not command-by-permission.* The fact that dumping of decision load is often necessary to avoid self-reinforcing degradation provides insight and support to the intuition that command-by-negation is advantageous as compared to command-by-permission. In the command-by-permission protocol, a lower-level decision-making component detects a condition, formulates a plan for action, sends a request for permission to execute the action to the higher-level component, waits for the permission (or denial) to arrive, and then executes the action. In the command-by-negation protocol, the lower level component does *not* wait for permission, but proceeds to execute the action when the time is right, while being prepared to abort the action if the higher-level component responds negatively. Clearly, avoidance of degradation by dumping at the higher-level component can be done more effectively in command-by-negation – if the higher-level component ignores the message, it does not prevent the lower-level component from executing the desired action. The same dumping of messages at the higher-level component in command-by-permission prevents the lower-level component from executing the necessary actions to exploit an opportunity or to block a threat. In both cases, dumping enables the decision-making organization to avoid the degradation, but in the case of command-by-permission this avoidance leads to greater rigidity and passivity.

*Minimize the need for coordination.* Minimizing coordination loops, both vertical and horizontal, reduces susceptibility to self-reinforcing degradation. Although reduction in coordination may appear counter-intuitive and controversial, some of the human factors literature has been calling attention to the potential negative impact of coordination requirements for a long time, e.g., Morgan and Bowers [15] cite findings from Naylor and Briggs [22] as follows: "...the performance of operators in a simulated air-intercept task was superior when the subjects worked independently of one another. Decrements in performance were observed when operators were placed in an organizational structure that encouraged interaction among the operators." Experimental findings (e.g., [13]) show that teams tend to perform better when they are able to communicate less under high-stress conditions. In the design of decision-making organization, assigning tasks to minimize the need for coordination reduces the amount of knowledge the team members need to have about each other's roles, and the amount they need to communicate, which can result in better overall performance [14]. An adversary, however, can compel increased coordination requirements by injecting misleading or conflicting information, or by presenting an organization with events that, by their very nature, require highly coordinated responses from multiple sub-organizations.

*Insulate the weak link.* Weaker decision-making components within the organization can accelerate the collapse of the entire system. It is advisable to insulate such a component from the rest of the system either by providing a greater degree of supervision or, if unavoidable, by allowing such a component to fail in its mission without expending excessive effort on the part of the superior component. This, however, may not be possible when the weak component is engaged in a critical task. This is also a situation that an intelligent adversary will attempt to create and exploit.

*Diagnose on-line and compensate by dynamic reorganization.* In several experiments, we observed consistent symptoms of the onset of collapse manifesting themselves well in advance of the actual collapse. This observation suggests a possibility of introducing an on-line diagnostic mechanism. Further, diagnosis leads to a possible compensation by dynamic reorganization, such as we explored in this paper (Figures 8 and 9). From the perspective of an adversary, however, the diagnostic and reorganization functions constitute excellent high value targets.

Now, in spite of this rich harvest of practical insights, we must remind ourselves that intellective models are not intended to produce definitive recommendations. Their value is in generating useful suggestions: suggestions for high-fidelity modeling approaches, suggestions for the design of experiments and suggestions for possible policy or structure redesigns. All these are to be explored and validated by other more concrete models or experiments. Despite their limitations, the strength of intellective models is that they are uniquely capable of identifying a phenomenon that would be difficult to detect in a pure and readily recognizable form in a real organization or even in an emulative model. By avoiding unnecessary detail, intellective models offer tools that help researchers visualize and explore the dependencies and the impacts of key factors in a manner that is relatively easy to grasp.